%% file: main.tex
\documentclass[10pt,conference]{IEEEtran}
\usepackage{cite}
\usepackage{amsmath,amssymb,amsfonts}
\usepackage{algorithmic}
\usepackage{graphicx}
\usepackage{textcomp}
\usepackage{xcolor}
\def\BibTeX{{\rm B\kern-.05em{\sc i\kern-.025em b}\kern-.08em
    T\kern-.1667em\lower.7ex\hbox{E}\kern-.125emX}}

\usepackage{enumitem}
\usepackage{pifont}
\usepackage{hyperref}
\usepackage{svg}

\newcommand\fakeSection[2][1]{
\ifthenelse{#1=0}{
\noindent\textbf{#2}
}{
\smallskip \noindent\textbf{#2}
}
}

\newcommand\blfootnote[1]{%
  \begingroup
  \renewcommand\thefootnote{}\footnote{#1}%
  \addtocounter{footnote}{-1}%
  \endgroup
}

\newcommand{\track}[1]{{#1}}

\newcommand\PAGANINI{\textit{PAGAnInI}}

\begin{document}

\title{Walking Down the Road to Independent Mobility: An Adaptive Route Training System for the Cognitively Impaired}


\author{\IEEEauthorblockN{Konstantin Rink\IEEEauthorrefmark{1}, 
Tristan Gruschka\IEEEauthorrefmark{1}, 
Patrick Palsbröker\IEEEauthorrefmark{1}, 
Marcos Baez\IEEEauthorrefmark{1}, 
Dominic Becking\IEEEauthorrefmark{1},
Udo Seelmeyer\IEEEauthorrefmark{1}, \\
Gudrun Dobslaw\IEEEauthorrefmark{1} and 
Patricia Stolz\IEEEauthorrefmark{1}
} \IEEEauthorblockA{\IEEEauthorrefmark{1}Bielefeld University of Applied Sciences, Bielefeld, Germany\\
Email: \{name.surname\}@fh-bielefeld.de}
}

\maketitle

\begin{abstract}
In this paper we describe the design and development of a \textit{route training system} for individuals with cognitive impairments  (CIs) living in residential care facilities.  Learning to move autonomously in public spaces is a fundamental skill for people with CI, who face several challenges to independently and safely move around. Yet, exploring opportunities for route training support, especially in residential settings, has received very little attention.
To explore these opportunities, we followed a design and development process based on inclusive design practices that considered the organisational context and aimed at involving people with CI in the software design. To ensure our solution addressed the identified needs and abilities of this heterogeneous population, we further framed the route training definition as a \textit{design process} that is enacted by the system, making the trainer and user co-creators of a personalised training. In this paper we report on the needs and challenges for mobility training in residential settings, introduce the design and formative evaluation of the route training system, to conclude with reflections and considerations on our methodological approach.
\end{abstract}

\begin{IEEEkeywords}
route training, cognitive impairments, inclusive design
\end{IEEEkeywords}

%

\noindent\textbf{General Abstract}. Learning to navigate public spaces without assistance is important for people with cognitive impairments (CIs). It can help them overcome challenges to independently and safely reach places in their daily lives. Yet, the use of technology for route learning has not been fully explored in research, especially for people with CIs living in residential care.
In this article, we describe the process of developing a \textit{route training system} that explores the use of technology support. In our research, we tried different ways to involve people with CIs in the design of the system, which is seen as a more inclusive approach to designing solutions.
We identified that people with CIs have different needs and abilities when it comes to route learning and related skills. For this reason, our solution focuses on ways to personalise the training, making sure people with CIs are involved in the personalisation, so that it fits their needs, abilities and learning progress.   
 %
\blfootnote{This is a post-peer-review, pre-copyedit version of an article accepted to the ``Software Engineering in Society" (SEIS) track of the 45th Internationl Conference on Software Engineering, ICSE 2023.}

\input{sections/introduction-short}
\input{sections/related-short}
\input{sections/methodology-short}
\input{sections/requirements}
\input{sections/design-short}
\input{sections/evaluation}

\input{sections/discussion}

\section*{Acknowledgement}


This work was supported by the German Federal Ministry of Education and Research under the funding program ``Forschung an Fachhochschulen" No 13FH514SX7, as part of the  PAGAnInI (`Personalized Augmented Guidance for the Autonomy of People with Intellectual Impairments") project.

The authors would like to extend their gratitude to Marleen Bönhoff for her contribution in providing illustrations for this paper.

\bibliographystyle{IEEEtran}
\bibliography{references}

\end{document}

%% file: sections/introduction-short.tex
\section{Introduction}

The ability to move independently is a critical skill for people with cognitive impairments (CIs)~\cite{wolf2021mobilitatsbildung}. It can be a catalyst for performing daily living activities, staying physically active, developing social relationships and having a higher quality of life \cite{slevin1998independent}. 
However, people with CI face several challenges to independently and safely move around, such as difficulties associated with \textit{cognitive processes}, e.g., learning the required mobility skills, spatial navigation, \textit{accessibility of navigation aids}, e.g.,  understanding maps and other navigation abstractions \cite{slevin1998independent}, and \textit{environmental factors}, such as institutional and structural barriers~\cite{slevin1998independent}.
%
These barriers limit their autonomy, which can lead to  sedentary behavior \cite{duggan2008impact}, have a negative impact in their psychosocial wellbeing  and general health~\cite{teipel2016information}. Mobility difficulties also make it unsafe to move around, which can lead situations such as injuries or people getting lost \cite{bantry2015dementia}.

\textit{Orientation and mobility} programs aim at training individuals in concepts, required skills and techniques to integrate sensory information, spatial orientation and movement to enable safe and independent mobility~\cite{chang2020orientation}. 
Aspects commonly addressed in mobility training with people with CIs include pedestrian education, use of public transportation and route learning~\cite{wolf2021mobilitatsbildung,brown2011designing,freina2015mobile,covaci2015assessing}.
This practice is commonplace with individual with vision impairments but less widespread and formalised among individuals with CIs~\cite{wolf2021mobilitatsbildung}. 
This is further evidenced, as we will see, by the few works on mobility training support in the literature.  
Instead, the main focus of research has been in exploring specialised \textit{navigation} solutions (like Google Maps but) for people with CIs, i.e., the use of technology support to provide instructions on how to reach a destination~\cite{rassmus2014finding,poppinga2014navmem,ramos2013geo}.
However, for a population with limited navigation skills, training is the recommended approach,  to not further suppress the development of such skills~\cite{brown2011designing,oliver2008learning}.


\begin{figure*}[ht!]
  \centering
  \includegraphics[width=\linewidth]{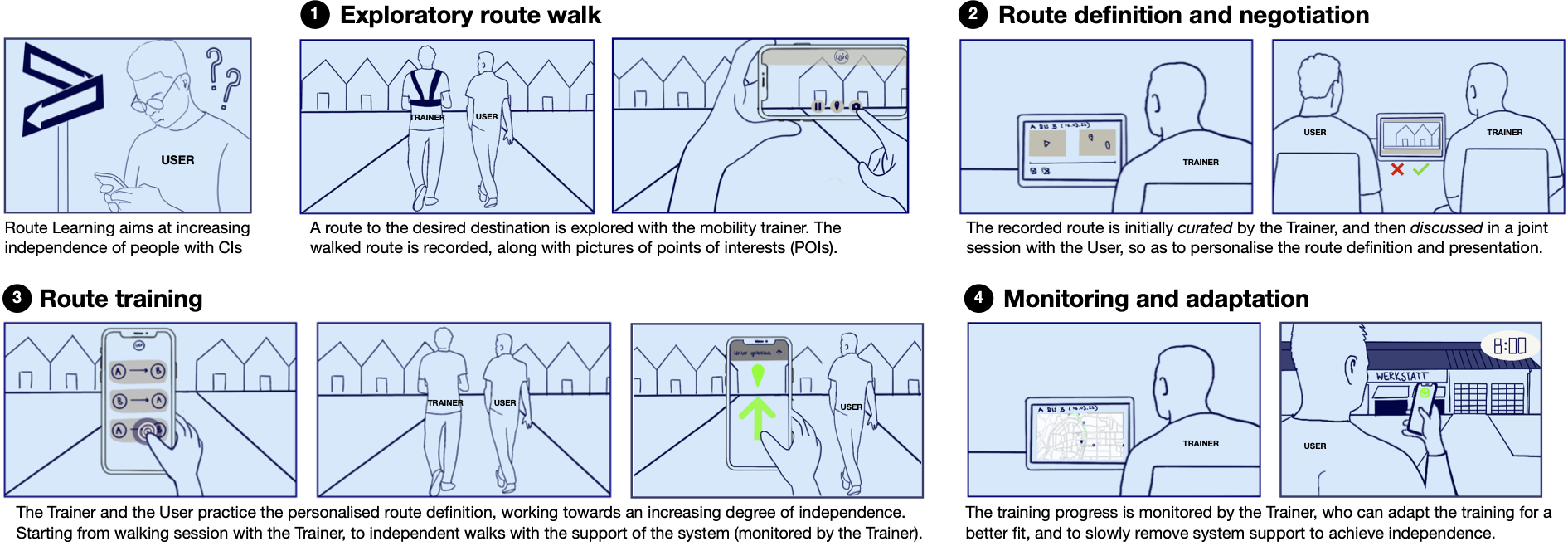}

  \caption{Route training training process that emerged from our Human-Centered agile process.}
  \label{fig:storyboard}
\end{figure*}

In this work we describe the \textit{design} and \textit{development} process of a \textit{route training system} for people with CIs (\track{see Figure~\ref{fig:storyboard} for an overview}). We focus on the specific challenges and opportunities of delivering mobility training in \textit{residential settings}, i.e.,  residential communities designed to foster and provide services for individuals with disabilities after they leave their family of origin~\cite{seifert2016wohnen}.
This is an unexplored socio-technical setting in mobility training that requires special attention to existing organisational practices, resources and limitations to ensure software solutions can be effectively adopted by individuals and organisations.
Due to the heterogeneity of the population in this context, we do not take specific clinical categorisations of CIs but rather adopt a functional perspective (as in \cite{hu2015investigating}), i.e., we focus on individuals with CIs affecting their mobility.




We followed a human-centered agile development approach, adopting inclusive design practices to involve individuals with CIs in the design and engineering process \cite{baez2018agile,spencer2020including}.
Through extensive user research activities with two prominent residential facilities in Germany, we uncover the barriers, design challenges and requirements for mobility training in organisational settings. These informed the concept of a route training system, which was developed and evaluated by leveraging different methods (from interviews, to focus groups to technology probes) to allow for the most diverse sources of feedback.
The resulting system, namely \PAGANINI, aims at enabling individuals with CIs to follow a personalised route training program that can adapt to their goals, abilities and learning progress, while lowering the barriers to implementing route training in organisational settings. 
In this development process, we address the following research questions:


\begin{itemize}[leftmargin=13pt,topsep=5pt]
    \item What are the main barriers, practices and design challenges to introducing mobility training in residential settings?
    
    \item How do we design a mobility training system to address individual and organisational needs and challenges?
    
    \item What methodological aspects need to be considered when engineering mobility solutions for residential settings?
    
\end{itemize}

In doing so, (i) we contribute with general insights and requirements for route training, as they relate to both the system \textit{design} and \textit{deployment} in organisational settings. We (ii) designed and piloted a personalised route training system that frames personalisation as a design process enacted by the system, where trainer and trainee co-participate in adapting the route learning to the different goals and abilities of the trainee. (iii) Contribute with reflections on our methodological approach, and especially the involvement people with CIs in the design and development process.
Overall, our experience point to the feasibility of involving people with CIs in the software design, given the right methods, and the feasibility (and value) of encoding their involvement in the personalisation of services targeting them.
In what follows we present the related work, design and development activities that were performed to address the main questions.

%% file: sections/related-short.tex
\section{Related Work}

\fakeSection[0]{Wayfinding and navigation assistance.} 
Wayfinding and navigation technologies aim at assisting users in reaching their destination by relaying instructions. 
Here, the focus is in exploring navigation strategies and designs so that people can understand and follow instructions, not on learning how to reach their destination without support.
%
Three prominent types of mobility approaches have been evaluated with individuals with CIs. 
\textit{Directional awareness}, where the idea is to  aid basic orientation tasks through a digital compass. This approach was explored in the Home Compass app~\cite{rassmus2014finding}, which aimed at supporting older adults with memory declines to freely move in their neighborhoods while keeping a reference point to home. \textit{Landmark-based navigation} where the idea is to rely on landmarks along a route as primary means of providing direction and supporting decisions. This approach has been adopted in several systems, such as the NavMem system~\cite{poppinga2014navmem} supporting wayfinding for individual with memory declines.
\textit{Turn-based navigation}, that aim at providing specific turn-by-turn navigation instructions to users. This approach is implemented in general-purpose navigation systems (e.g., Google Maps), and adopted by some navigation systems for people with CIs, such as the handheld AR-based system by  Ramos et al. \cite{ramos2013geo} that provides turn-by-turn direction overlaid on the mobile phone camera.
Previous studies (e.g., \cite{gomez2015adapted}) 
point to the landmark-based navigation as the most effective, less cognitively demanding and preferred approach for individuals with CIs. 
%
%
While these systems do not support training, we take inspiration from the navigation strategies to inform design decisions of the wayfinding component of our solution.

\fakeSection[0]{Mobility training for people with CIs.}
The design and development of mobility training support for individuals with CIs has been investigated in three prominent research projects.

RouteMate~\cite{brown2011designing} is a route learning system for people with intellectual disabilities developed by the RECALL project. The approach is based on scaffolding the learning of new routes, i.e., going through different levels of support  until reaching independent mobility. A mobile app for the users offers training of \textit{map-based concepts} in three different levels or modes of support. The \textit{planning} mode, where a route (start, destination) is defined with the help of a trainer, and augmented with points of interest (POIs) enriched with text, pictures and symbols. The \textit{practice} mode allows the user to rehearse the prepared route, and offers location-based reminders when getting close to POIs, and alarms when deviating from the defined routes. In the \textit{use} mode the user is encouraged to use their own skills through the gamification of the route navigation. The screen turns off while navigating between POIs, and when reaching one, users are challenged to select the picture that corresponds to the POI. 
This system was then further developed in the direction of game-based learning \cite{brown2013engaging}. The use mode was replaced with a \textit{challenge} mode, which incorporated digital scavenger hunts (e.g., a ninja-game route) aimed at reinforcing map concepts and route learning.
%
While this project produced valuable contributions, (i) its navigation is based on maps, which is known to be pose challenges to people with CIs \cite{gomez2015adapted}. (ii) The scaffolding is limited to three  modes that apply to the entire training, preventing the it from adapting more closely to the trainee (e.g., learning progress) and to the route (e.g., parts of the route being more familiar or difficult). (iii) The solutions do not account for how the training can be personalised.


In the same spirit, the SmartAngel \cite{freina2015mobile} project aims at enabling people with intellectual impairments, and particularly individuals with Down Syndrome, to move around independently in a urban context. The project developed solutions providing support in two steps: a \textit{pre-training} phase where navigation-related skills are trained in a simulated environment and a second phase, where  \textit{on-the-street} monitoring is provided for actual route navigation. The pre-training phase apps aim at addressing the longer training time required due to learning challenges by reducing the on-the-street training. Here, FriendlyRoad \cite{freina2014learning} provides a virtual environment where the user can safely  train mobility skills associated to street dangers before moving onto actual streets. 
%
The on-the-street support \cite{freina2015social} provided is essentially a monitoring system. It targets individuals with CIs that already have the ability to move independently, monitoring their walks in real time and acting as a safeguard so that tutors can intervene when needed. 
As we can see, these solutions are targeted at (i) educating people on road safety rather than route training, and (ii) monitoring for people already able to move  independently, instead of providing on-the-street training.

The POSEIDON project developed solutions for training and navigation support for individuals with Down Syndrome. A \textit{stationary system} \cite{covaci2015assessing} provided a VR environment, where scenarios could be systematically manipulated for individualised training. In this system, a virtual trip is customised and enhanced by a caregiver, with landmarks, POIs and dangerous points. The trainees would then learn and rehearse the route in the virtual environment. A \textit{mobile app} \cite{kramer2015developing} would then support the actual navigation on the streets. This app displays the route definition of the stationary system, relying on map representations and picture aids. Caregivers can monitor trainees but not interfere with the navigation. Results from pilots indicated problems reading the map, interpreting instructions, and fulfilling user needs with off-the-shelf directional assistance. 
As seen, these solutions focus on pre-training instead of actual on-the-street training, and are limited by the representation adopted (maps). They also do not address the personalisation of the training to the individuals ability and progress.

Overall, these few projects and solutions \track{(including others focused on public trasportation \cite{carmien2005socio})} explored interesting perspectives of mobility training for people with CIs. However, important aspects such as \textit{personalisation} of the training, and strategies to support the \textit{scaffolding} of the learning, are largely underdeveloped. The challenges and requirements of introducing these solutions in \textit{institutionalised settings} is also not addressed. In our work, we address these gaps to contribute with a (i) better understanding of mobility challenges in residential settings, (ii) an approach to personalisation that relies on making trainers and trainee co-participants in the design of the training, and a  (iii) learning scaffolding approach that relies on the adaptation of the degree of support and supervision, aided by learning indicators.









%% file: sections/methodology-short.tex
\section{Methodology}


 We followed a \textit{human-centered} agile development approach, adopting strategies from the literature to integrate design and development to engineer solutions for vulnerable populations~\cite{baez2018agile,brhel2015exploring,carroll2016aligning}. In terms of the process, we integrated user research activities as a pre-requirement phase (e.g., as seen in \cite{carroll2016aligning}) to get an understanding of the users and the complexity of the problem space before diving into the design and development. 
 \track{After this preliminary phase, we run parallel design and development interwoven tracks, allowing design cycles at least one sprint ahead to inform subsequent developments~\cite{brhel2015exploring}. 
 These incremental cycles advanced the design and development through three main conceptual phases,}
 from (1) \textit{concept development}, aiming at shaping the overall focus and concept of route training support to (2) \textit{prototyping the experience} of using the system, involving users in the foreseen training process  to (3) \textit{producing design and software artifacts} of different level of maturity.
We summarise these phases below.


\fakeSection{Pre-requirements phase.} We first approached interest groups and organisations with the purpose of getting a better understanding of the entire context and establishing partnerships with organisations interested in participating in the project. We partnered with two main organisations in Germany, which gave us access to residential facilities and sheltered workshops for people with disabilities. We performed \textit{ethnographic observations}, held meetings with different stakeholders to help us establish a working approach to the collaboration.
Next, we carried out semi-structured \textit{interviews} with individuals with CIs as well as professionals (e.g., rehabilitation managers, caregivers), in order to get insights into the barriers, needs and opportunities for mobility training in the target scenario. 

\fakeSection{Concept development.} What emerged from the previous phase are insights and requirements that shaped the concept of a route training system. We iteratively refined \textit{personas} and \textit{storyboards}, which help us illustrate the concept putting representative users in specific target scenarios. In parallel, this informed the technical specifications of the system, components and constraints.
The resulting concept was then discussed with individuals with CIs and professionals in \textit{concept validation} activities, assessing the main components of the solution in scenario-based interviews driven by storyboards.

\fakeSection[0]{Experience prototyping.} We prototyped the main route training process by relying on \textit{technology probes}~\cite{hutchinson2003technology} early on, and partial  prototypes of the system in later iterations. The idea was to immerse the users in the experience of the training, as intended according to our route training concept. This emerged as a necessary and effective way of eliciting specific feedback from our target users.

\fakeSection[0]{Developing software artefacts.} 
With a well defined design space, the focus turned into design and software artefacts of higher maturity. From a design perspective, this was driven by feedback sessions with prototypes of varying level of detail, from paper to high-fidelity prototypes, focusing on:   navigation \textit{metaphors} (e.g., symbols, icons), \textit{accessibility} features (e.g., colors, fonts), and \textit{organisation} and usability of system components. The development followed up on the footstep of the design (two sprints behind) to engineer the required support.

\smallskip

\noindent In the above process, the type of activities and the of level involvement was tuned to allow for a wide rage of feedback from people with CIs, which is emerging as an important aspect of \textbf{inclusive design practices} \cite{spencer2020including}. 
The process was overseen by an Ethics Review Board, and assisted by the experts in the partner organisations.

In this paper we report on the early stages of the design and engineering, from the identification of user needs and elicitation of requirements, to the initial development and validation of the concepts and the approach to personalisation.

%% file: sections/requirements.tex
\section{Understanding Users and Context }

 We initially explored the barriers, needs and opportunities for supporting people with CIs in ``route training", studying the specific context of residential communities of people with disabilities. 
 In the following we describe the methods, insights and resulting requirements from the comprising activities.

\subsection{Methods}
\label{sec:context-methods}
The pre-requirements phase involved  visits and interviews that led to the definition of high-level requirements for mobility training support.\footnote{Study materials can be found at \url{https://bit.ly/3TaWuC8}} 
In the \textit{preliminary visits}, we engaged with two main organisations in Rhine Westfalia, Germany, and visited 5 of their facilities to perform ethnographic field observations, and informal introductory discussions with different stakeholders. We documented detailed visit protocols, including notes regarding observations and discussions.
\textit{Guided interviews} were conducted with (1) individuals with CIs to gain basic knowledge about mobility practices and personal goals. The study participants (N=7) consisted of people with different intellectual backgrounds. The age range was 26-54. We also conducted interviews with (2) care professionals (N=10) of different roles to provide us the organisational perspective, and act as proxies to elicit further input about mobility challenges and practices. The interview guides are provided as supplementary materials.
 
The collected data was analysed using qualitative content analysis~\cite{ritchie2013qualitative}. 
We first transcribed the recorded interviews, organised by guiding question. From these, we identified relevant passages and organised them in themes with fixed definitions. This process was first done by independently by two researchers and then combined via consensus. In what follows we discussed the emerging themes from the relevant dimensions of our analysis.

\subsection{Results}

\fakeSection[0]{Characterising the context.} 
We explore a common framework for supporting people with disabilities in Germany. The people we focus on live in disability care facilities  and work at sheltered workshops. We describe the purpose of these  below.

\textit{Institutional disability care}. The German system of care for people with disabilities has been undergoing a profound transformation since the 1990s.
Nowadays, people with disabilities have a broad spectrum of options from \textit{institutional living} to \textit{outpatient assisted living} communities to \textit{outpatient support} in one's own home~\cite{seifert2016wohnen}. 
People with intellectual disabilities can decide for themselves where they want to live after leaving their families. In Germany,  60\% of them are currently cared for in \textit{inpatient institutions} after leaving their parental home, while just under 40\% make use of outpatient support~\cite{seifert2016wohnen}. We focus on inpatient institutions in this project, which is a large but often forgotten scenario in technology support.  These institutions provide a communal setting that provides care services and aim at supporting the integration of residents into society.
Services include assistance with daily activities, such as personal hygiene, meals, laundry or accompaniment to recreational activities. A caregiver is present around the clock, so that staff is on site at night if there are problems.

\textit{Workshops for people with disabilities} (short: WfbM). These are special vocational institutions for people with disabilities. They come into play when all other measures to promote the labour market participation of people with disabilities have been unsuccessful~\cite{schreiner2017teilhabe}. 
Around 280,000 workers are employed in these workshops, partly as full-time, partly as part-time workers. In Germany, they are \textit{``legally designed as rehabilitation facilities with the  aim of placing suitable employees from the workshop in the general labour market"}~\cite{schreiner2017teilhabe}. In certain locations, the workshops are also responsible for the commute to the workshop. 
If the employees are not able to get to the workshop on their own, the workshop must provide a transport service for them. The transport services are either private companies, such as bus companies, or they are charities that have their own driving services. 
The transport pick people up from their homes and take them directly to the workshops. 

\fakeSection{Barriers to independent mobility.}
The interviews revealed different types of barriers to independent mobility in our target scenario.
Some individuals, due to their condition, simply lack the \textit{cognitive} or \textit{motor} skills to move on their own, and require to be accompanied or driven around. When such conditions are not an issue, individuals might still lack proper \textit{training} in road safety, proper reading, ability to understand timetables and clocks and other important skills for independent mobility. 
In terms of \textit{organisational} barriers, organisations might not have the personnel and resources for training, or might feature rigid schedules that discourage individuals from choosing independent means of transportation. The \textit{environment} can also limit mobility, e.g., with certain routes being demanding or not safe for walking, limited availability of public transportation for the route, or changes in the environment (e.g., snow or darkness). \textit{Social} factors were also mentioned, starting from over-protective parents to hostilities faced in public transportation (e.g., comments and looks from school students). Motivational and \textit{emotional} factors also play a role, as some individuals who would be able to move independently do not dare to do so (e.g., they are afraid, panic, get frustrated) or find the driving service just too convenient. 
As a consequence of the above, individuals in this context generally rely on the driving service, which is essential for them, but also one that gets in the way of having a more independent and active life.

\fakeSection{Mobility goals.}
Personal mobility goals gain special importance as an incentive in route learning. 
Both long and short distance goals aimed at overcoming boundaries to meet private needs on their own and independently: Visiting relatives and friends, visiting cities, shopping privately (rather than in an externally determined group). 
Underlying these goals, personal expectations were highlighted, from just being able to make it to arriving the fastest way possible.  
Although a large proportion of respondents had low levels of technology sovereignty (e.g., not all of them had access to smartphones and the internet), clear wishes were expressed for technology-based mobility training. Simple instructions, voice output and an emergency button made it clear that the technical solution should fulfil a broad spectrum of interaction and comprehension abilities.

\fakeSection{Mobility training and practices.}
The interviews and visits revealed that, currently, mobility training is not formally integrated in organisational practices. It is generally provided by professionals in the residences as part of supporting the life goals of the residents. Given a life goal, as small as being able to go for shopping alone, professionals would assess whether the goal could be achieved and determine what resources are available to fulfill this goal. It was stressed, however, that there are limited resources in the residential facilities and that route training is a ``time puller".  Underneath formal and informal mobility training activities, however, we identified a set of valuable practices that facilitate route training:


\begin{itemize}[leftmargin=10pt,topsep=5pt]
    \item \textit{Assessing the feasibility} of independent mobility. Professionals use their expertise and contextual knowledge to assess the feasibility of independent mobility for a route.
    
    \item \textit{Supporting the development of required skills}. Facilities provide basic courses (e.g., use of mobile phones) that support required skills for (tech-supported) mobility training.
    
    \item \textit{Breaking down mobility goals}. Some professionals will set smaller learning goals before the main goal. e.g., learning the route to a nearby shop before shopping downtown.
    
    \item \textit{Learning by doing}. Professionals train the residents along the actual routes so that they learn by actually walking them.  
    
    \item \textit{Adapting the training}. Professionals stressed the wide spectrum of needs and the different initial situations of individuals. This requires them to personalise their approach to the specifics of the individual and ``make progress with them".
    
    \item \textit{Incrementally pulling out support}. Professionals scaffold the training support, initially leading the route walks to then test the learning (i.e., quiz individuals at decision points) and eventually letting  trainees take over.
    
    \item \textit{Setting up emergency lines}. Emergency lines are available in case residents need help or get lost. Residents have these numbers in their phones and can request help when needed.
\end{itemize}


In terms of technology, there is no solution currently in place to support mobility training. Professionals stressed the importance of training and that the individuals \textit{``learn and not just follow instructions"}, pointing out that solutions like Google Maps as not being appropriate in terms of design and focus. And while they supported the idea of technology for mobility training, they were concerned that mobile phones could be a distraction and a risk to road safety. Organisationally, they expressed that keeping track of the whereabouts of the residents would facilitate digitisation of the training but were hesitant due to the implications for privacy.

\subsection{Requirements for the system}
The pre-requirement phase revealed general needs, practices and opportunities that should shape the design of the route training system. 
We summarise the main requirements below:

\begin{itemize}[leftmargin=17pt,topsep=5pt]
    
    \item [R1] \textit{Keep the expert in the loop}. Professionals have accumulated knowledge and expertise, so it is crucial to support them (and their practices) in the training rather than replacing them, as well as the personal attention. 
    
    \item [R2] \textit{Allow for the adaptation of the training to the individual goals and progress}. It is fundamental to account for the different learning pace, background, and goals, and scaffold the training based on their learning progress.

    \item [R3] \textit{Account for people's different abilities to interpret instructions}. Here, the system should adapt to how guidance is relayed to make guiding and training users more effective. 

    \item [R4] \textit{Make the system accessible and usable}. Differences in interaction abilities should be considered to design an accessible system that can be  used by a wider set of users.     
    
    \item [R5] \textit{Make the mobility support safe to use}. The solution should consider safety of use by design, minimising the attention required by the system and avoiding distractions in traffic. 
    
    \item [R6] \textit{Address negative feelings} associated with the training. Feelings of uncertainty and fear of getting lost, should be reduced by  features that address them at the source. 
    
    \item [R7] \textit{Incorporate safety nets in case of unexpected situations}. Provide mechanisms to recover from unexpected situations arising from human, technology and environment. 
    
    \item [R8] \textit{Provide transparency, protection and control over user privacy and data}, especially when monitoring. Users should be made aware of the implications of data sharing, its purpose, and provided with control mechanisms. 

    \item [R9] \textit{Incorporate strategies to scale the training process} in organisational setting where there is a limited number of trainers. As expressed by professionals, technology  should ``make their work easier, not make them work more".

\end{itemize}






In general terms, the above points to the need for \textit{scaling} supervised training practices (R1, R9), \textit{personalising} the route training to relevant dimensions (R2-R4), and enabling \textit{barrier-free training} support on-the-street (R5-R7), while being mindful of \textit{privacy} and data protection aspects (R8).

%% file: sections/design-short.tex
\begin{figure*}[ht!]
  \centering
  \includegraphics[width=\linewidth]{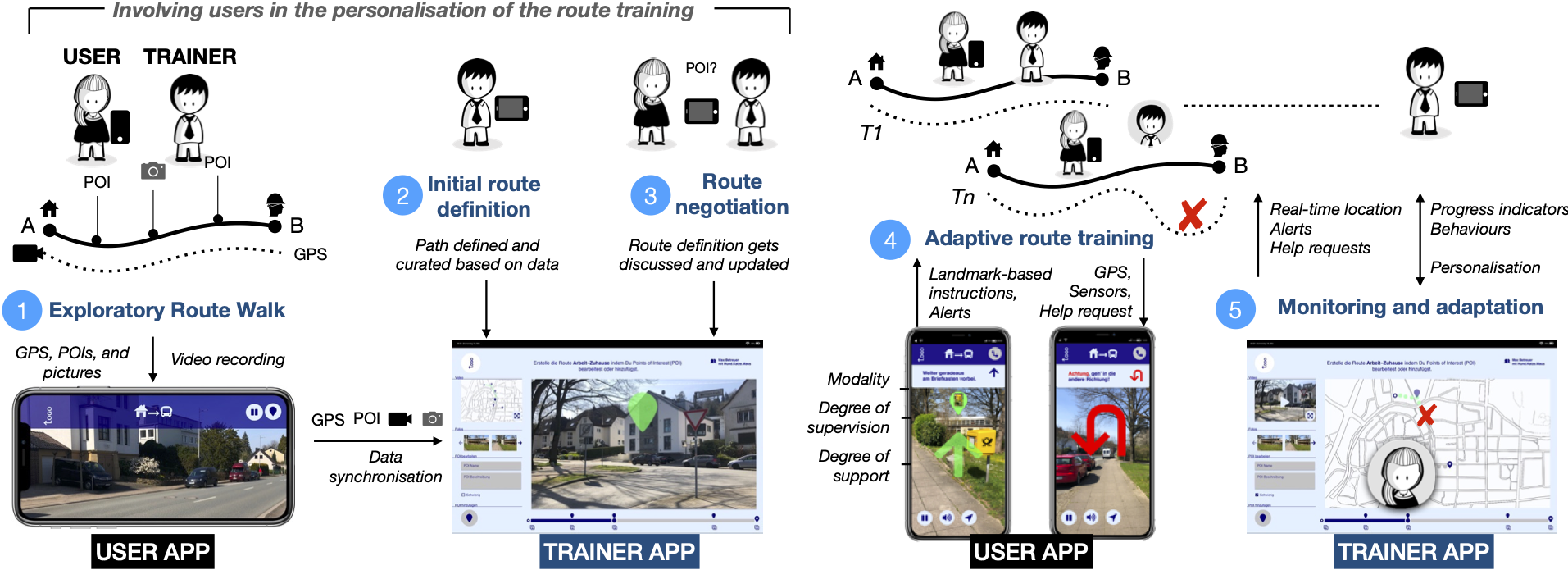}

  \caption{General training process illustrating the interaction points of the main actors with the system components}
  \label{fig:architecture}
\end{figure*}

\section{System Design and Rationale}


The initial phase led to the design and development of a \textit{route training system}, namely \PAGANINI, to address the mobility needs and barriers in our target context. 
Figure \ref{fig:architecture} illustrates how the users interact with the two main apps of  the \PAGANINI ~system to carry out the route training. \track{The system addresses the main requirements with the following general properties:} 
%

\begin{itemize}[leftmargin=10pt,topsep=5pt]

\item \track{\textit{Personalised training} (R1-R4). The system aims at enabling individuals with CIs to follow a \textit{personalised} route training program tailored to their goals, abilities and preferences, under the supervision of a mobility trainer. The personalisation is achieved by enacting a process that involves both the individual and the trainer in the definition, planning and presentation of the training. }

\item \track{\textit{Barrier-free training support} (R5-R7). The actual route training relies on experiential learning, supporting \textit{on-the-street training} with features that address actual and perceived \textit{barriers} to route training.  We provide reassurances during navigation to reduce negative feelings, detection and recovery from unexpected situations such as going off-track, and the ability to request help when needed.}

\item \track{\textit{Adaptation to learning progress} (R1, R2). The system allows mobility trainers to monitor and adapt the training (along certain adaptation dimensions) based on the learning progress of the individuals.} 

\item \track{\textit{Scaffolding the training support} (R1, R9). We scaffold the involvement of trainers, going incrementally from human supervision to monitoring to system-based supervision in order to \textit{scale} the training and advance the learning goals at the individual's pace.}

\end{itemize}

%
\track{The system is the result of design, prototyping and testing cycles that led to its development and refinement. 
In the following, we describe in detail the concepts and latest design, and in the next section the insights from our validation efforts.}

\subsection{Route training concepts}

\fakeSection[0]{\track{Target users.}}
We will denote simply as \textit{user}, to the individual with CIs that acts as trainee, and \textit{trainer} to the care professional that acts as mobility trainer. \track{We iterated over various definitions of representative \textit{personas} (not described here) and arrived to a set of characteristics and assumptions for target users that fits the organisational practice.}
For the \textit{users}, we focus on individuals with the motor and visual perception abilities to walk without assistance. 
These additional barriers would have required a differential mobility training approach.
We also assume that users went through basic smartphone training, which is indeed available at the partner organisations, and through pedestrian education. 
For the \textit{trainer}, we consider professionals with mobility training education, and basic technology skills. We assume there will be training in the organisations to prepare trainers in technology-based route training.






\fakeSection[0]{Concepts.}
\textit{Route training} is an important aspect of mobility training that focuses on training people in finding their way from a place A to B (e.g., home to work). People with CIs are challenged in their ability to generalise and transfer knowledge, so wayfinding requires rehearsing and learning each specific route instead of just  general navigation strategies~\cite{freina2014learning,covaci2015assessing}. 
Next, we introduce relevant route training concepts:

\begin{itemize}[leftmargin=10pt,topsep=5pt]

\item \textit{Way}. A way is the (street-)connection a user wants to be able to walk independently, e.g., to get home from work. 
Ways are not bidirectional, as the impressions, scenery and instructions change with the direction of walk. 

\item \textit{Path}. One exact (street-)path to walk along in a specific way. A way can contain several possible paths that differ in distance, time, complexity, traffic situation, detours, etc.

\item \textit{Route}.
A path augmented with landmarks and reassurances that serves as the basis for route training.

\item \textit{Point of interest (POI)}. Places that stand out to the user during a route walk, which could be adopted as landmarks or reassurances in the training. They are described by a coordinate, timestamp, and at least one photo. 

\item \textit{Landmark}. Landmarks are objects that are not subject to rapid change and may have a personal reference for the user (houses, trees), where a decision must be made. 

\item \textit{Reassurance}.  Reassurances are prominent points along a route (e.g., church tower, house, tree) where no decision must be made, but which encourage the user to be on the correct path and helps to remember the next landmark.

\end{itemize}

The above is the basis for \textit{landmark-based} wayfinding. We adopted this approach as it has been shown to be better suited for individuals with CIs (e.g., \cite{poppinga2014navmem}). In this approach, the  idea is to rely on \textit{landmarks} along a route as primary means of providing direction, rather than turn-by-turn directions. 



\subsection{Route training system}

We now turn to the process depicted in Fig.~\ref{fig:architecture}. We highlight the design space, design choices and ingredients of the system,  as resulted from our iterative process. Note that we describe the current system specification and not the implementation.

\fakeSection{Exploratory route walk.}
The first step in the training process is the \textit{exploratory route walk} (ERW). This is a preliminary joint walking session driven trainer along the intended route, with the purpose of i) assessing the abilities of the user and the feasibility of walking independently along the route by leveraging his expertise and contextual knowledge, and ii) collecting and eliciting relevant data for the definition of the route training.  This activity is supported by the ERW component of the User mobile app. 

The walking session starts at the beginning of the route (e.g., Home) and ends at its destination (e.g., Work). The specific path to follow between these two points is decided by the trainer based on the specific abilities, background and goals of the user (e.g., path featuring memorable characteristics, simpler or safer to navigate, faster). 
Indeed, the trainer may explore and define different routes. 

During the walk, the system collects information necessary for the training. This includes i) \textit{GPS points} recorded along the way so as to reconstruct the path actually walked, ii) candidate \textit{POIs} with pictures and additional notes, necessary to support landmark-based navigation, and iii) \textit{video recording} of the walk to facilitate the analysis for the route definition and negotiation. GPS data and video are recorded automatically, but the POIs should be elicited by the trainer based on their experience and best practices. 
As sensitive data is recorded during this process (video and pictures), all the ERW data is stored on the local mobile phone of the user and transferred to the Trainer app through a peer to peer connection. This data is curated by the trainer in the next step of the process (route definition), and synced to the cloud only then. Some data such as video recording is never synced to the server. 








\fakeSection{Route definition and negotiation.}
The next step in the process is the \textit{route definition}. Here, the trainer analyses all the information collected in the ERW to curate a ``prototype" route definition that can be later discussed with the users. Unlike previous approaches that rely on off-the-shelf navigation instructions or automatically generated landmarks \cite{kramer2015developing,ramos2013geo}, we rely on the experience of the trainer, and user involvement, to define a personalised route. This activity is supported by the Route Design component of the Trainer app.

To support the analysis and reflection on the collected data, this component allows the trainer to playback the entire ERW session. This is done by synchronising video playback, the walking progress on a map representation of the route, with markers on playback progress and the map, signaling geotagged POI and pictures. With this visualisation, the user can not only replay the session, but also jump to any POI along the way to watch / listen to the discussion around it. 
The Trainer can then curate the route definition. This implies adding, editing and removing POIs and pictures, curating instructions and adding notes for the discussion with the user. The path can also be edited, e.g., to fix potential GPS issues. 


Next, the prototype route definition is jointly discussed with the user in what we call the ``negotiation", to derive a working route definition that fits the individual. The \textit{negotiation} aims at involving the user in the definition of the route, extending the idea of inclusive design practices from software alone to the personalisation of the training. Thus, there are two main goals to the negotiation: making sure (i) we select appropriate route elements (e.g., landmarks) to maximise \textit{learning} (e.g., is this a potentially effective landmark?), and (ii) the user would be able to follow the instructions, so as to maximise \textit{comprehension}  (e.g., are these instructions understandable?).

The system facilitates the route negotiation through a simplified version of the Route Design interface, that uses a slideshow metaphor to go over POIs one by one, emphasising pictures and instructions (instead of a map, which proved ineffective in our experience prototyping). Here, the trainer would elicit the required feedback on each POI, pictures and instruction, to then apply or annotate the feedback.








\fakeSection{Route training.} 
The \textit{route training} is the actual learning activity, where the trainer and the user practice the negotiated route definition, working towards the increasing degree of independence of the user. 
 With the User app, the user follows the route definition from start to destination with a given \textit{degree of supervision} (e.g., assisted by an in-person human trainer). Wayfinding instructions are relied with a specific \textit{degree of support} (e.g., showing explicit landmark-based instructions) and displayed to the user in specific \textit{modalities} (e.g., text and audio). Upon unexpected situations detected or self reported (e.g., user lost, panic), the system can prompt recovery options to assist the user. While practicing the route, the system collects relevant data to compute progress indicators and enable monitoring and adaptation. 
A described here, our \textit{adaptive route training} approach thus considers three dimensions for adapting the training: i) degree of supervision, ii) degree of support, and iii) interaction modality, which we describe next.  
 
%



First, the supervision provided by the trainer goes from in-person support and remote training support to system-based support. The starting point is \textit{in-person} supervision, where trainer assists the user in the landmark-based navigation of the route, using their mobility training expertise to increase the independence until it is safe for the user to practice the route on their own. Next, the trainer can provide a similar assistance via a \textit{remote} audio link supported via a monitoring interface. In the last degree of supervision, the training \textit{app takes over} the on-the-road training, with the trainer taking a passive role in monitoring the progress.

The second dimension to the adaptation is the degree of wayfinding support. We rely on strategies that incrementally aim at removing the assistance provided by the app. \textit{Actionable mode} relay explicit guidance within the landmark-based navigation, notifying users of decision points and reassurances. The \textit{quiz mode}, adopted from the observed mobility training practices, prompts users about the way to go when reaching decision points. The quiz acts as a checkpoint that encourage attention and provide valuable signals to analyse the progress of the user. The \textit{reward mode} turns off instructions at decision points to provide retrospective feedback, rewarding users for making the correct decisions or alerting them of mistakes made past the decision points. Finally, \textit{mute mode} turns off the assistance on a POI. It is important to note that these strategies can be applied to specific sub-paths along the way, and not only as a general strategy for the general route navigation (contrary to \cite{brown2013engaging}). In our experience prototyping, this emerged as an  aspect to address, since users have might experience different level of familiarity or difficulty along the route.

The third dimension of the adaptation is the instruction modality. Given the different interaction and comprehension abilities, the interface rely on multiple representations including text, symbols (e.g., arrows, signals), audio, tactile and augmented reality (AR), which can be made available to the user under certain constraints for safety of use. 
%
AR  can only act as a supplementary modality. When in doubt, the user will activate the AR modality -- by lifting the phone and pointing the phone camera on the road -- and will get in-place directions over the real world as seen through the phone camera. While this constant flow of instructions is very useful, limiting its use minimises potential risks in outdoor environments~\cite{rovira2020guidance}.


In terms of data collection, the training component collects information about user location, application usage, and unexpected situations in order to build performance indicators. These signals are also shared in real time with the monitoring component to enable remote assistance. For transparency, the user is informed before each practice session of the information collected, and is requested to grant permissions. Whether to allow users to disable tracking is still a contested features as we will see in the preliminary evaluation.

\fakeSection{Monitoring and adaptation} 
The training process comes full circle with the \textit{monitoring} of the user progress. The main goal here is to enable the (i) remote monitoring and assistance of users during the route training, and (ii) facilitate the progress assessment and adaptation of the training for future training sessions. These activities are supported by the Trainer app.

A monitoring dashboard enables the remote supervision of users during their training activities. This includes an audio link and real time monitoring of the user location, along with indicators of unexpected situations. 
To support the assessment and adaptation the key is the definition of performance indicators that capture the self-reliance of the user on the road.  We are currently envisioning indicators that express \textit{autonomy}, inversely associated with the reliance on instructions and assistance; \textit{accuracy}, as the ability to make correct wayfinding decisions; \textit{errors}, capturing mistakes and unexpected situations on the road; \textit{recovery}, as the ability to self-recover from errors. We also aim at capturing self-reported indicators (e.g., confidence), as the perceived readiness can play a major role in the ability to move independently. 
These indicators should be aggregated to allow the trainer to reflect on the learning progress from different perspectives. It should allow trainers to observe the \textit{learning trend}, comparing the training progress over time. It should also describe the \textit{progress along the route},  allowing trainers to spot parts of the route where the user is making progress and where problems arise.

The trainer will then be able to perform an informed decision regarding whether to adapt the training along the three dimensions previously described. We are also exploring the possibility of having system recommendations with a human-in-the-loop approach. 
However, as the monitoring plays a supporting role, its design heavily depends on the design decisions of the other components. With the other components maturing, we are getting ready to prototype this activity.











%% file: sections/evaluation.tex
\section{Preliminary Evaluation}

The goal of the evaluation  was to obtain feedback to guide and validate design decisions in the development process. We summarise the feedback in (i) \textit{concept validation}, where we sought feedback on the general direction of the route training system concept, and (ii) \textit{experience prototyping}, where we sought feedback on the underlying training process. The focus at this stage was to assess the feasibility of our approach to involving people with CIs in personalising the training. Detailed study materials are available online.\footnote{Study materials can be found at \url{https://bit.ly/3TaWuC8}}

\subsection{Concept validation}

\fakeSection{Methods.}
The concept validation considered two aspects: (i) feedback on the concept of the route training system, and (ii) the implications of its implementation in the organisational practice.
%
To obtain feedback on the \textit{concept},  we initially conducted scenario-based interviews with the help of a (printed) storyboard. We engaged individuals with CIs (N=2) as well as care professionals (N=2).
As we target an organisational setting, we also examined the implications of deployment for the organisation. To this end, we combined the data collected in the pre-requirement phase, as well as the concept validation interviews to distill the main implications by performing qualitative content analysis (as in Section \ref{sec:context-methods}).


\fakeSection{Feedback on the concept.}
From the user perspective, the storyboard was effective in communicating the concept. The users expressed that such a solution would be useful, but could not anticipate what could be missing or provide suggestions. 

On the other hand, professionals expressed that, if successful, PAGAnInI would bring more\textit{ mobility options} to the residents, and enable their \textit{independent mobility}. They mentioned that, currently, there is a pressure towards the driving service, which reduces their self-determination. From an organisational perspective, the  benefit of system  would be in \textit{supporting the mission} of the institution. However, as such practice is not currently widespread, they expressed concerns about the readiness of the organisation and the impact on the overall organisational practice, which we discuss in detail next.







\fakeSection{Impact on organisational practice.}
Participants were quick to stress that a system alone cannot address mobility training issues, and the whole deployment context should be considered. Among the main aspects to consider they mentioned:

\begin{itemize}[leftmargin=10pt,topsep=5pt]
    \item \textit{Defining new processes and protocols}. The organisation would need to define new processes to integrate mobility training in their practice, and define clear protocols for how to handle various unexpected situations emerging from it.
    \item \textit{Defining roles and delegating responsibilities}.  The training will require the definition of new roles and tasks that need to be defined and delegated within (and across) organisations.
    \item \textit{Adapting roster structure}. The roster structure needs to account for the new tasks related to the training, in terms of working hours and availability of personnel.
    \item \textit{Implementing technology training}. Introductory courses on \PAGANINI ~and related technologies should be provided, for users and  personnel, since not all are tech-savvy. 
    \item \textit{Coordinating with driving services}. If more people are able to walk independently, people would need to be signed off from the driving service, either temporally or seasonally. 
    Potential disruptions to the service need to be factored in.
    \item \textit{Adopting more flexible schedules for residents}. Letting people practice their routes might affect time-sensitive commitments, e.g., getting to work. These commitments should be negotiated to offer residents a real mobility alternative.
    \item \textit{Determining legal repercussions}. Participants expressed uncertainty about the legal repercussions of introducing the system, given that the training will be mediated by technology and part of the assistance would be provided remotely.
    \item \textit{Sorting out ethical aspects}. Ethical concerns about the  monitoring of residents were raised. Mechanisms for providing consent and data protection were deemed important. 
\end{itemize}

Despite the above concerns, participants expressed their support for the concept, and reiterated that the goals of the intended solution aligns with those of the organisations.



\subsection{Experience prototyping}

\fakeSection[0]{Methods.}
The experience prototyping took place in several steps, covering the main steps of the training process. For the (1) \textit{exploratory route walk}, we initially prototyped the experience by relying on the phone camera as a technology probe~\cite{hutchinson2003technology}. 
A researcher took the role of trainer, who drove the discussion with users (N=12) around potential POIs in ``walking interviews".
On a second step, a functional prototype was used under the same study settings. 
%
%

The (2) \textit{route negotiation} activity was prototyped to assess the feasibility of involving users in the route training definition, explore representations that could facilitate the activity, and characterise the feedback that could be obtained. To do so, we took the data collected in the previous ERW sessions and iterated over two representations: (i) \textit{maps} (highlighting the walked path) and (ii) \textit{photo grid} (emphasizing pictures).  In the latter, we arranged the pictures as a table in a word processor, with rows denoting different POIs in sequential order and columns the pictures taken on each POI. With these, a researcher would then drive the negotiation activity by discussing on the route definition with the users (N=9). 

The (3) \textit{route training} was prototyped to elicit further feedback by making the user experience the training activity. Three training sessions were organised, and driven by a care professional so as give us a window into the actual practice. 

In all three prototyping experiences, there was at least one researcher acting as a passive observer, taking notes on the experience (behaviors and interaction with the technology). These observations were qualitatively analysed to characterise the user involvement in the activity, and design insights.







\fakeSection{Exploratory route walks.}
Prototyping this activity opened up concrete discussions about the possibilities of the application (especially the working prototype). When showing the creation of a route the mobile app, with the types of locations for start and destination (displayed as icons), the users voiced \textit{concrete mobility goals} that were not captured in the initial interviews, especially for leisure. For example, one user described that she would love to go to the park during lunch breaks, and even suggested a flower icon for it.

In terms of  technology use, this activity was originally intended to be driven by the trainer, who would be operating the app to capture POIs and pictures. However, users showed interest in \textit{taking an active role} in capturing the POIs themselves and were able to do so. This opened up further opportunities to engage users in the process. 

Participants appreciated the \textit{transparency} feature,  informing them of the data that was being collected before the start of the ERW. However, in the elicitation and discussions around POIs, we identified \textit{latent privacy concerns} that could prevent a more tailored route definition for the users. 
During the ERW session, the researcher noticed that a landmark had caught the attention of the user (a cigarette dispenser) but was not brought up. In the debriefing, it turned out the user did not want the trainer to question their smoking habits.  
Potential trade-offs between personalisation and privacy need further investigation.

We also identified the importance of perspectives in pictures. Users were not able to recognise salient POIs (e.g., a church) from a different angle. This highlights the importance of \textit{personalising pictures} to their walking perspectives, and the potential issues of using stock pictures (e.g., Google Street View photos as in~\cite{gomez2015adapted}).






\fakeSection{Route negotiation.}
In terms of representations, we initially attempted to discuss the route definition by relying on a maps representation. This resulted to be too abstract as already pointed out in the literature (for navigation)~\cite{gomez2015adapted}, but ultimately the actual path resulted not to be central to eliciting feedback.
We then decided to emphasise pictures in a picture grid (as described before) and used this as the basis for the discussion, which unlike our first attempt was easy to follow.

In the discussions, users were able to \textit{engage and provide valuable input} in different forms. Users were able to discern what \textit{POIs were relevant} to them.
Along the same route, users had different relevant POIs, depending on their background, e.g., a user that was passionate about nature, selected nature-related landmarks. This highlights the need for \textit{personalised} POIs as opposed to general landmarks.
Users could also \textit{identify bad pictures}, i.e., pictures they could not recognise, because the main feature they associated to the POI was not captured.
Indeed, having multiple pictures taken on the same POI provided \textit{options to choose from}, and the representation facilitated deciding on the one that evoked the most recognition. These experience support our personalisation approach, giving a voice to people with CIs in the route training.








\fakeSection{Route training}
Involving users in route training sessions guided by a mobility trainer provided interesting design insights. 
We observed different dynamics between trainer and user at different parts of the route. Users took a passive role following the trainer in unfamiliar or more difficult-to-navigate areas. In contrast, when reaching \textit{islands of familiarity}, the user would take over and lead the way with more confidence. What this tells us is that users might feature different navigation abilities (or learn at a different pace) at different parts of the route. This requires a finer level of granularity for scaffolding the training and support than the one provided by the state of the art (e.g., \cite{brown2011designing}), allowing \textit{adaptation at sub-paths} rather than at the whole route level.



We also observed \textit{varying levels of engagement} and attention, with a higher degree at places that evoked personal interest. While in this context, the trainer could intervene and try to involve the user, in the technology-assisted training the system should implement strategies to keep the user in check. Our approach, as described in the design section, capture some of these strategies to rely instructions (e.g., quiz mode) and implement safety measures.
Further pilots are necessary, however, to assess the effectiveness of these (and other) strategies in keeping users focused and engaged.


Debriefing users on the activity, we identified aspects related to mobility goals and privacy. Users expressed that they would take different routes to a destination depending on contextual factors (e.g., the longer route on a sunny day, or a shortcut when in a hurry), which is addressed in our system by allowing multiple route definitions per way.
On the matter of tracking during training to facilitate monitoring, users did not express any privacy concerns. They indeed contested the idea of having a ``turn off tracking" feature suggested by the care professionals, as they value their safety over other aspects.




%% file: sections/discussion.tex
\section{Discussion}
The design and preliminary evaluation of the system uncovered important aspects for route training in residential settings. We identified important requirements that directly affected the system design.
%
A key aspect uncovered was the importance of personalisation, and the main dimensions that should be considered when personalisating technology-mediated training support, including goals and abilities, personal relation to the routes, and learning pace. In realising the personalisation, we devised an inclusive approach that sought to involve people with CIs in the route training definition. Our experience prototyping activities indicate that such involvement is both feasible and valuable in implementing personalisation. Furthermore, we identified important implications for system deployment and organisational practices, most of which go beyond system design.
Addressing these requires a concerted effort with the organisations to undertake the larger socio-technical context.

Throughout the design and development process, we also learned valuable lessons in regards to our methodological approach, which we summarise below.





\fakeSection{Involving people with CI in the design and development process.} 
The methods we have resorted to in our research and technology development were oriented towards inclusive design practices~\cite{spencer2020including}, breaking stereotypes to consider the basic skills that an autonomous and sovereign self would have. We encountered however some challenges in involving our target group, with methods achieving varying levels of effectiveness.
\textit{User interviews} are useful means of eliciting input, 
but they are more effective when interviewed persons have a receptive and expressive vocabulary that enables them to \textit{``understand the questions and formulate an answer that sufficiently reflects their own perspective"}~\cite{niediek2015nicht}. The majority of our user interviews were limited in these abilities, and despite efforts in simplifying the activity, the insights were limited without concrete artefats. 
\textit{Focus groups} have the advantage that many perspectives can be collected in a relatively short time. These activities were more lively in our experience, but we faced a similar challenge in that participants who did not have the semantic and pragmatic abilities disappeared into passivity. 
In the initial iterations, having access to other \textit{stakeholders as proxies} to get further insights resulted valuable in getting a more informed perspective. 
We started to engage participant more actively when \textit{prototyping experiences} with users. Be it with cultural probes or partial prototypes, we were able to observe participants behaviors (noting aspects they are often not able to verbally express), participants were able to provide more concrete  feedback (agreeing and disagreeing with design choices), and even contribute with original ideas (e.g., types of destinations to be considered and how to represent them).
%
The above insights contribute to the empirical experiences in involving people with CI emerging in recent literature~\cite{spencer2020including}.

\fakeSection{Involving users beyond software design, and into the co-creation of artefacts served by the software.}
 Involving individuals with CI in the design and development process can be an effective approach to the produce solutions that fit their needs and abilities. We argue however, that the involvement should be factored in not only in the software design but also in the production of artefacts offered by software systems. In particular, we noted that in order to serve the training needs of individuals, a design process needed to be enacted by the system to produce a route training definition that would adapt to the diverse needs and backgrounds of individuals. That is, the training becomes a design artefact on its own, and the trainer and user become co-participants in a design process that is enacted after the deployment of the software. Encoding participant involvement in the software design can be an effective way of ensuring continuous participation beyond the development process.

\fakeSection{Leveraging the organisational context to address design challenges.}
An important aspect when designing for an organisational context is to consider how this context affects design decisions, i.e., How is this different from designing directly for an end-user? 
Organisations can introduce complexities in the design and deployment, but also opportunities for deploying a true socio-technical system that can orchestrate humans and technology to address design challenges. For as we observed in this project, technology alone cannot address all the issues but a concerted efforts between organisation, users and technology. 
In this project, for example, trainers can be considered experts in mobility practice, and training preparation activities can be assumed as imparted by the organisations. In contrast, addressing end-users directly would have required further encoding practices and training knowledge right into the system. What this means is that, as result of the design of the socio-technical system, software artefacts as well as recommendations for the development of an education curricula for trainers, and protocols for organisations, should be produced. This is an aspect we are actively working toward.

\section{Conclusion}
In this paper we uncovered important requirements for the design and deployment of route training support in residential settings. We identified the importance of personalisation, and dimensions that should be considered when personalising for technology-mediated learning support. The proposed route training system implements a personalisation approach that seeks to actively involve people with CIs in the definition of the training -- an approach that we has shown to be feasible and valuable in experience prototyping activities. 
In our design and development process, we also explored different methods to involve users in the software design. That involvement is feasible, supporting inclusive design practices, but we experienced higher success when prototyping experiences with users. 

As part of our ongoing efforts, we are iterating on higher maturity prototypes and planning further pilots. 

\fakeSection{Limitations.}
The evaluation is based studies with small numbers of participants. However, 
the focus was to obtain the most varied type of feedback through participant involvement, to guide the system design and assess the feasibility of our route training approach and methodology.